# Airy-function electron localization in the oxide superlattices

Z. S. Popovic* and S. Satpathy

*Department of Physics, University of Missouri, Columbia, MO 65211, USA*

Oxide superlattices and microstructures hold the promise for creating a new class of devices with unprecedented functionalities. Density-functional studies of the recently fabricated superlattices of lattice-matched perovskite titanates $(SrTiO_3)_n/(LaTiO_3)_m$ reveal a classic wedge-shaped potential originating from the Coulomb potential of a charged sheet of La atoms. The potential in turn confines the electrons in the vicinity of the sheet, leading to an Airy-function localization of the electron states. Magnetism is suppressed for structures with a single $LaTiO_3$ monolayer, while the bulk antiferromagnetism is recovered in the structures with a thicker $LaTiO_3$, with a narrow transition region separating the magnetic $LaTiO_3$ and the non-magnetic $SrTiO_3$.

PACS: 73.20.-r, 73.21.-b

While the growth of the atomically abrupt, lattice-matched interfaces between semiconductors has long been perfected and such structures are now widely used in electronic devices, the growth of such high-quality oxide structures has not been possible for a long time. In a recent work, Ohmoto et al.[1] reported the growth of atomically precise, lattice-matched superlattices made up of alternating layers of $LaTiO_3$ and $SrTiO_3$. What is striking is that the quality of the interfaces easily matched the quality of similar semiconductor structures. This has raised the hope of being able to grow quality structures using other oxides[2] as well, opening them up for new physics and potentially novel device concepts.

An understanding of the interface electron states played a key role in the development of semiconductor electronics and revealed many novel features not found in the bulk constituents, e. g., the formation of the two-dimensional electron gas and the subsequent discovery of the Quantum Hall effect. Analogously, new electronic properties may be anticipated for the oxide heterostructures, where strong correlations between electrons will undoubtedly lead to new physics, not found in the semiconductor counterparts. In this paper, we study the interface electronic structure of the recently fabricated perovskite titanate superlattices using calculations based on the first-principles density-functional theory. An interesting finding is the wedge-shaped potential well caused by the positive sheet-charge density at the interface and the Airy function localization of the electron states in the potential well. The Airy function localization has been observed[1] in the form of an exponential-like spread of the $Ti^{3+}$ fraction near the interface. To our knowledge, this is the first time that a wedge-shaped potential has been found in any system, which opens up the possibility for studying the physics of a new two-dimensional Airy gas.[3]

The bulk electronic structure of both the constituents of $(SrTiO_3)_n/(LaTiO_3)_m$, the subject of this paper, has been well studied. Both form in the perovskite structure, with a three-dimensional network of corner-sharing TiO$_6$ octahedra.[4,5] While SrTiO$_3$ is a band insulator with an empty Ti (d$^0$) bands, LaTiO$_3$, in contrast, is a Mott insulator with Ti (d$^1$) occupancy. In a simple picture, the latter can be thought of within the context of the half-filled Hubbard model, leading to an antiferromagnetic insulator as observed. However, the insulating state is very quickly destroyed with the introduction of a small concentration of holes via the addition of extra oxygen[6] or via Sr substitution (with as little as x ≈ 0.05 for La$_{1-x}$Sr$_x$TiO$_3$).[7] The quick destruction of the insulating state by the hole doping is reminiscent of the Nagaoka state where a single hole destroys both the antiferromagnetism and the insulating behavior in the half-filled Hubbard model.[8]

The behavior of the electrons in the superlattices such as (SrTiO$_3$)$_n$/(LaTiO$_3$)$_m$ is expected to be fundamentally different from that in the random alloy with the same composition. As opposed to some effective average potential seen by the conduction electrons (the stripped-off La (d$^1$) electrons) in the alloy, the potential in the heterostructure is due to a coherent superposition of the Coulomb potentials of the La ions organized on lattice planes. The La (d) electrons become detached from the sheet but become confined in the wedge-shaped Coulomb potential of the sheet charge. Strong screening due to the embedding SrTiO$_3$ material weakens the Coulomb potential, spreading thereby the bound electron into several layers of SrTiO$_3$ on either side of the interface. A Hartree-Fock model study by Okamoto and Millis[9] has shown that this delocalization of the electron state leads to a metallic behavior.

To study the electronic structure of the (SrTiO$_3$)$_n$/(LaTiO$_3$)$_m$ superlattices, we have performed density-functional calculations within the local-spin-density approximation (LSDA) as well as the "LSDA+U" method.[10] Since the LSDA+U method is necessary to describe the correlation physics for bulk LaTiO$_3$, albeit in an approximate way,[11,12] the majority of calculations reported here are performed with this method. The linear muffin-tin orbitals (LMTO) method[13] is used throughout to solve the Kohn-Sham equations. For superlattices with a thick LaTiO$_3$ part, we have used the full octahedral distortions of bulk LaTiO$_3$, while for superlattices with a single LaTiO$_3$ layer, we have used undistorted octahedral structure of the embedding SrTiO$_3$.

Consider the (SrTiO$_3$)$_n$/(LaTiO$_3$)$_1$ superlattice grown along the (100) direction as in the experiment. Here a single LaTiO$_3$ layer is embedded in a thick SrTiO$_3$ part and with the La (d$^1$) electrons stripped from it, the layer behaves like a two-dimensional sheet of positive charge. The Coulomb potential of the sheet charge is determined using elementary electrostatics, where according to the Gauss' law, the electric field is given by $E = \sigma/2\varepsilon_0$, where σ = 1 |e|/S is the sheet charge density and S is the interface surface area per La atom. The field is reduced due to screening and leads to a wedge-shaped potential with the minimum located at the plane of the sheet charge. This potential in turn binds the La (d$^1$) electron in its Coulomb field.

The variation of the potential seen by the electron in the solid may be calculated from the variation of some reference energy in the density-functional calculation. A convenient reference energy that we have successfully used for the semiconductor interfaces is the cell-averaged point-charge Coulomb potential $V$.[14] It may be calculated by first averaging the potential parallel to the plane near the interface (planar-averaged potential) and then by averaging over a period normal to the plane. We can, alternatively, calculate V by first averaging over the volume of the Wigner-Seitz atomic spheres:

$$V_i = \frac{3q_i}{2s_i} + \sum_j^{\prime} \frac{q_j}{|r_i - r_j|}, \qquad (1)$$

and then by averaging over all spheres with a weight factor proportional to their volumes:[15] $V = \sum_i \Omega_i V_i / \sum_i \Omega_i$, where $\Omega_i = 4\pi s_i^3/3$ is the sphere volume, $s_i$ is the sphere radius, $r_i$ is the sphere position, and $q_i$ is the total charge, nuclear plus electronic, and $i$ is the sphere index. In Eq. (1), the first term is the sphere average of the potential of the point charge located at the center of the muffin-tin sphere and the second term is the Madelung potential due to all other spheres in the solid.

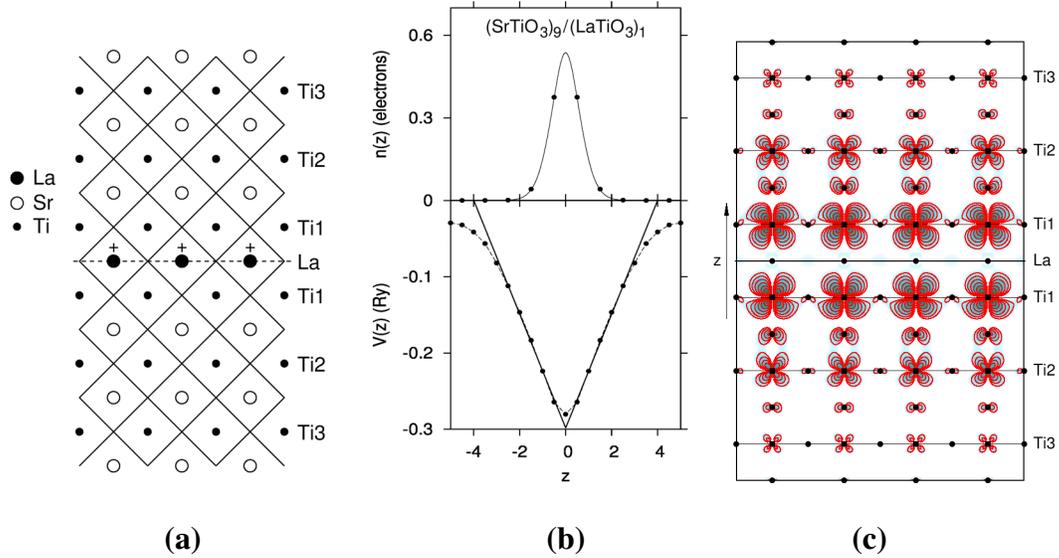

(a)  (b)  (c)

**Figure 1** (color online) The $(SrTiO_3)_n/(LaTiO_3)_1$ superlattice structure (a) with a monolayer of $LaTiO_3$ embedded in bulk $SrTiO_3$. The positively charged La sheet produces a wedge-shaped potential (b) following Gauss' law in elementary electrostatics. In the bottom part of (b), dots denote the cell-averaged electrostatic potential V(z) calculated from Eq. (1) with distance z from the interface in units of the $SrTiO_3$ monolayer thickness, while the solid line is a fit corresponding to a uniformly charged sheet. The potential confines the stripped-off La ($d^1$) electrons in the interface region (c). The contours show the electron charge-density, integrated between the conduction bottom and the Fermi energy, indicating the electron leakage into about three Ti layers on either side of the interface. Contour values are: $\rho_n = \rho_0 \, 10^{n\delta}$, where $\rho_0 = 3.4 \times 10^{-3} \, e/\text{Å}^3$, $\delta = 0.31$, and $n$ labels the contours. The top part of (b) shows the electron density n(z), which is integrated over a perovskite layer at the position z, with the solid line being a fit using the lowest-energy Airy state. All results were calculated with the local spin-density approximation in the density-functional theory.

The cell-averaged potential V(z) as a function of the distance z from the interface is shown in Fig. 1. The potential has a remarkable, text-book like wedge shape, V = -eE|z|, as expected for the constant electric field near a sheet of charge, leading to a novel wedge-shaped quantum well. If we compare the computed electric field with that in the vacuum electric field, E = σ/2ε$_0$, we infer a large dielectric constant of ε ≈ 23. It is well known that SrTiO$_3$ is close to being ferroelectric with an unusually large long-wavelength dielectric response of ε ≈ 30 x 10$^3$ at low temperatures.[16,17] However, the screening of the sheet charge involves the short-range dielectric response, with a smaller expected ε, consistent with the calculated electric field. The truly wedge-shaped potential occurs for a theoretical sheet charge of zero thickness. The finite thickness, of one atomic layer in the present case, causes the bottom of the potential to flatten out at the center of the sheet, a feature that is present in our calculated potential. Indeed, the inversion symmetry dictates that the electric field is strictly zero at the center of the sheet.

The stripped-off La(d$^1$) electrons which produce the positive La sheet charge in the first place, become bound in turn in the wedge-shaped potential well. The eigenstates of the electrons are obtained from the Schrödinger equation,

$$\frac{-\hbar^2}{2m}\frac{d^2\Psi}{dz^2} + V(z)\Psi = E\Psi, \qquad (2)$$

where V(z) = F|z| is the potential of the sheet charge and F = -eE is a positive constant. Defining the scaled length $l = (\hbar^2/2mF)^{1/3}$, energy $\lambda = (2m/\hbar^2 F^2)^{1/3} E$, and coordinate $\xi = z/l - \lambda$, the Schrödinger equation takes the form $d^2\Psi/d\xi^2 - \xi\Psi = 0$, whose solutions are the well known Airy functions Ai(ξ). The complete solutions are obtained by joining the Airy functions at the interface (z = 0) and then by matching the function or its derivative: Ai(-λ)=0 or $dAi(\xi)/d\xi |_{\xi=-\lambda} = 0$, for the odd and even parity states, respectively. The lowest three eigenstates are shown in Fig. 2.

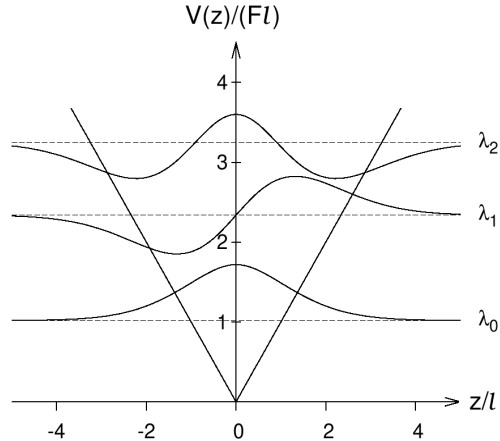

**Figure 2** Lowest eigenstates of the electron in the wedge-shaped potential well of a uniform sheet-charge density, with length and energies in scaled units (see text).

The lowest of the Airy states is occupied in the solid, resulting in a rapid decay of the charge of the electron away from the LaTiO$_3$ layer, as may be seen from the densities-of-states (Fig. 3). The electronic charge is predominantly Ti (d) like, located mostly on the Ti1 atoms and decaying rapidly as one moves away from the sheet (see also Figs. 1(b) and 1(c)). The calculated core level energies systematically track the wedge-shaped potential as well.

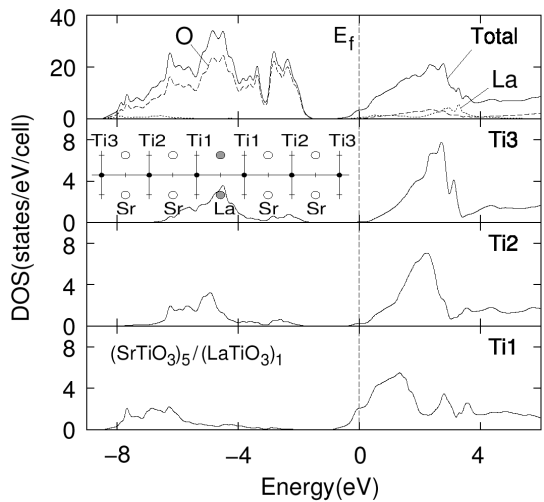

**Figure 3** Energy- and atom-resolved densities-of-states, indicating the rapid decay of the Ti conduction charge (mostly Ti (d)) as one moves away from the La sheet. In fact, very little conduction charge (conduction DOS integrated up to E$_f$) is visible from the figure for Ti atoms beyond the first layer.

The number of electrons occupying different layers may be obtained by integrating the density-of-states (DOS) from the conduction bottom to the Fermi energy for each perovskite layer as a function of its distance from the La sheet. The layer charges, so calculated, are shown by circles in the top part of Fig. 1, and they fit quite well to the shape of the lowest-energy Airy function (solid line). The length scale of the Airy function obtained from this fit is $l$=3.0 Å in excellent agreement with the expected value $l = (\hbar^2/2mF)^{1/3}$= 2.4 Å. Since the ground-state Airy function spreads to about ~3 $l$ on either side of the potential well, the electron wave function spreads to two to three layers of SrTiO$_3$ (layer thickness ≈ 3.91 Å). This is clearly seen from the charge-density contour plot (Fig. 1). The spread is in excellent agreement with the same obtained from the experimental EELS (electron energy loss spectroscopy) profile of Ti$^{3+}$ after deconvoluting it using the EELS profile for La. The latter serves as an effective resolution function,[18] as suggested by a comparison between the EELS data and the annular-dark-field image for La.[1]

The Fermi surface is shown in Fig. 4, which consists of five interpenetrating bands, the wave function characters of which are indicated in the lower right corner of the figure for the Γ point in the Brillouin zone. The orbital components of these levels are easily understood using insights from band calculations for the bulk materials.[12,19,20] The octahedral crystal field splits t$_{2g}$ below e$_g$ for Ti, while for La, e$_g$ is below t$_{2g}$. Furthermore, since the d orbitals of La have somewhat higher energy than Ti, it is clear that the Ti1(t$_{2g}$) electrons should have the lowest energy on account of their placement near the bottom of the wedge-shaped potential well. Of these, the two $d_{xy}$ states, corresponding to the two Ti1 layers on either side of the La sheet, have the lowest energy. Next come the doubly-degenerate Ti1 $d_{yz}$ and $d_{zx}$ orbitals, four in total, which are split into bonding and anti-bonding states, the latter occurring above the Fermi energy $E_f$. The next one up, the highest of the five states below $E_f$, is the $La(z^2-1)$ state, again, helped by its location at the bottom of the potential well. Above that is the $La(x^2-y^2)$ state and so on.

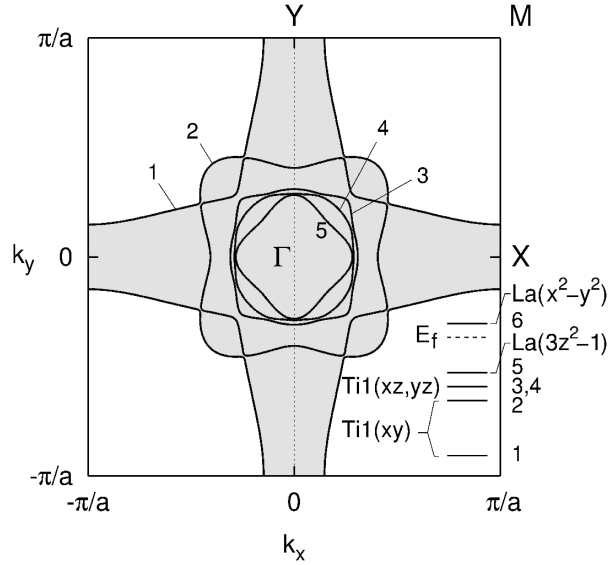

**Figure 4** The Fermi surface of $(SrTiO_3)_5/(LaTiO_3)_1$ shown in the surface Brillouin zone of the superlattice and its wave function character indicated in the lower right part of the figure.

The magnetic behavior of the superlattices show some interesting features as well. For the single $LaTiO_3$ layer structures (m=1), we find the paramagnetic state to be stable (both in the LSDA and the LSDA+U calculations), rather than the AF state of the bulk $LaTiO_3$. The paramagnetism of these superlattices is understandable in light of the fact that bulk $LaTiO_3$ becomes paramagnetic with a small amount of hole doping[6,7] and the leakage of the electron into the $SrTiO_3$ part dilutes the electron density and is tantamount to hole doping in the $LaTiO_3$ part. Along the same lines, we may reason that for a sufficiently thick $LaTiO_3$ layer, the inner part would be antiferromagnetic (bulk-like behaviour), while the outer Ti layers would have either reduced moments or may even become paramagnetic.

To test these ideas, we have performed an LSDA+U calculation for the $(SrTiO_3)_5/(LaTiO_3)_6$ superlattice structure, which has a six-monolayer thick $LaTiO_3$ part, taking the Coulomb and exchange parameters to be U = 5 eV and J = 0.64 eV, respectively, following earlier authors.[12] The results show the $LaTiO_3$ part to be antiferromagnetic, with the Ti magnetic moments equal to the theoretical LDA bulk value (~0.9 $\mu_B$). In addition, the two adjacent Ti layers at the interface with the $SrTiO_3$ region acquire a sizeable magnetic moment (about half of the bulk value) due to their proximity to the magnetic region. The magnetic moments of the rest of the Ti layers in the $SrTiO_3$ region are zero.

In conclusion, we have shown the formation of a quantum well with a text-book-like wedge-shaped potential originating from a uniformly charged $LaTiO_3$ sheet in the $(SrTiO_3)_n/(LaTiO_3)_1$ superlattices. The electronic structure is explained in terms of the Airy function localization of the electrons leaking out of the $LaTiO_3$ layer, but which become bound in the potential well. With a $LaTiO_3$ region of several layers thick, the inner layers show bulk-like antiferromagnetic behaviour, while the magnetism disappears in the embedding $SrTiO_3$ layers with a narrow transition region between them.

Acknowledgment. This work was supported by the U. S. Department of Energy under Grant No. DE-FG02-00ER45818.